# Low Thermal Conductivity and Interface Thermal Conductance in SnS$_2$

Saheb Karak[1], Jayanta Bera[2], Suvodeep Paul[1], Satyajit Sahu[2], and Surajit Saha[1*]

[1]Department of Physics, Indian Institute of Science Education and Research, Bhopal, 462066, India

[2]Department of Physics, Indian Institute of Technology Jodhpur, Jodhpur 342037, India

*Correspondence: surajit@iiserb.ac.in

**Abstract:**

After the discovery of graphene, there have been tremendous efforts in exploring various layered two-dimensional (2D) materials for their potential applications in electronics, optoelectronics, as well as energy conversion and storage. One of such 2D materials, SnS$_2$, which is earth abundant, low in toxicity, and cost effective, has been reported to show a high on/off current ratio, fast photodetection, and high optical absorption, thus making this material promising for device applications. Further, a few recent theoretical reports predict high electrical conductivity and Seebeck coefficient in its bulk counterparts. However, the thermal properties of SnS$_2$ have not yet been properly explored, which are important to materialize many of its potential applications. Here, we report the thermal properties of SnS$_2$ measured using the optothermal method and supported by density functional theory (DFT) calculations. Our experiments suggest very low in-plane lattice thermal conductivity ($\kappa$ = 3.20 ± 0.57 W m$^{-1}$ K$^{-1}$) and cross-plane interfacial thermal conductance per unit area (g = 0.53 ± 0.09 MW m$^{-2}$ K$^{-1}$) for monolayer SnS$_2$ supported on a SiO$_2$/Si substrate. The thermal properties show a dependence on the thickness of the SnS$_2$ flake. Based on the findings of our DFT calculations, the very low value of the lattice thermal conductivity can be attributed to low group velocity, a shorter lifetime of the phonons, and strong anharmonicity in the crystal. Materials with low thermal conductivity are important for thermoelectric applications as the thermoelectric power coefficient goes inversely with the thermal conductivity.

**Introduction**

Two-dimensional (2D) semiconducting materials beyond graphene have received tremendous interest because they have the required band gap suitable for applications in various fields, including photodetectors [1–4], light-emitting diodes [5,6], solar cells [7,8], biomedical sensors [9,10], and also thermoelectric power generation [11]. One of such 2D materials is the 2H phase of $SnS_2$, which is earth abundant, low in toxicity, and stable in atmospheric conditions [12–14]. A monolayer of 2H-$SnS_2$ consists of a close-packed layer of tin atoms sandwiched between two layers of sulfur atoms. The structure of 2H-$SnS_2$ is identical to that of 1T-$MoS_2$ [15–18], which is another promising 2D material with numerous potential applications. The very weak van der Waals force between the adjacent layers makes it easy to exfoliate down to monolayers with a relatively large area. However, contrary to $MoS_2$, upon thinning down to monolayer, the band gap of 2H-$SnS_2$ remains indirect [18]. Importantly, 2H-$SnS_2$ shows high values of on/off current ratio in field-effect transistor devices [19], high optical absorptions [20], and indirect band gap irrespective of the layer number even down to monolayer, thus making $SnS_2$ one of the most suitable materials for various applications.

Recently 2H-$SnS_2$ has seen extensive research activities to address its potential applications. However, its thermal properties have not been well explored so far. There are a few reports [11,21–24] that predict useful thermal properties of 2H-$SnS_2$, nonetheless, the experimental evidences are still minimal. Performance of thermoelectric materials is defined by the well-known figure of merit ZT= $S^2\sigma T/\kappa$, where S is the Seebeck coefficient, σ is the electrical conductivity, T is the temperature and κ is the total thermal conductivity due to electrons and phonons. A recent report by Lee *et al.* [21] has shown a high Seebeck coefficient and electrical conductivity as well as very low thermal conductivity in $SnS_2$ where they have investigated thick (16 and 21 nm) flakes of $SnS_2$. A theoretical study by Wang *et al.* predicted a very low in-plane thermal conductivity of ~ 8 W m$^{-1}$ K$^{-1}$ for monolayer 2H-$SnS_2$ [11]. Skelton *et al.* attributed the low thermal conductivity to intrinsic anharmonicity and low dimensionality and further predicted the out-of-plane thermal conductivity to be lower because of the weak interlayer interaction in the c-direction [22]. However, there is no experimental work, to the best of our knowledge, reporting the in-plane and out-of-plane thermal conductivity as a function of layer thickness and verifying these theoretical predictions. The experimental demonstration by Lee *et al.* [21] reveal a thermal conductivity of 3.45 W m$^{-1}$ K$^{-1}$ for ≈16-nm-thick flakes of $SnS_2$; however, it did not report the thermal conductivity for thinner layers down to monolayer as well as the interfacial thermal conductance.

Raman spectroscopy is a nondestructive characterizing tool that is very sensitive to the flake thickness of 2D materials, strain effects, defects, etc. Additionally, Raman spectroscopy has been established as a powerful tool to investigate the thermal properties of 2D materials by probing the thermal behavior of phonons [25]. This is of particular importance in the case of atomically thin 2D flakes, where a direct measurement of the thermal properties is a challenging task. Though the thermal conductivity constitutes both electronic and phononic contributions, the phononic contribution is of primary importance in the case of semiconducting materials like 2H- $SnS_2$. There has been extensive research in this field over the recent past and Raman-based measurements have been rigorously used to investigate the thermal properties of various 2D materials like graphene [25–27], h-BN [26,28–30], and transition metal di-chalcogenides (TMDs) [31–33]. In contrast to the case of a suspended device where the heat flows predominantly in-plane, in supported devices an additional channel for dissipation of heat is introduced through the interface with the substrate, neglecting the convection and radiation losses. To understand the thermal properties of any 2D material supported on a substrate, it is important to measure both the interfacial thermal conductance per unit area (g) and the in-plane thermal conductivity, simultaneously, which is very crucial for its potential electronic and thermoelectric device applications as well as fundamental interests. In our paper, we have carried out a systematic investigation of the thickness-dependent thermal properties of $SnS_2$ in terms of in-plane thermal conductivity as well as the interfacial thermal conductance through the interface of the thin layer supported on the standard $SiO_2$/Si substrate. We have prepared flakes of monolayer (1L), bilayer (2L), and five layer (5L) $SnS_2$, by mechanical exfoliation followed by dry transfer onto Si substrates coated with ~300 nm of $SiO_2$. We have observed a systematic increase in the interfacial thermal conductance per unit area (g) from $0.53 \pm 0.09$ to $3.60 \pm 0.65$ MW m$^{-2}$ K$^{-1}$ and an increase in the in-plane thermal conductivity (κ) from $3.20 \pm 0.57$ to $5.00 \pm 0.90$ W m$^{-1}$ K$^{-1}$ with an increase in the flake thickness from monolayer to five-layers. The κ of $SnS_2$ supported on $SiO_2$/Si is found to be orders of magnitude lower than other 2D materials like graphene (600 W m$^{-1}$ K$^{-1}$) [26,27], h-BN (280 W m$^{-1}$ K$^{-1}$), [26,30], $MoS_2$ (62.2 W m$^{-1}$ K$^{-1}$) [31], $MoSe_2$ (59 W m$^{-1}$ K$^{-1}$) [32], and $WS_2$ (32 W m$^{-1}$ K$^{-1}$) [33] thus making $SnS_2$ a potential thermoelectric material. To better understand the origin of low lattice thermal conductivity in the monolayer of $SnS_2$, we have performed theoretical calculations using density functional theory (DFT). Our calculated value of κ for monolayer $SnS_2$ is 4.92 W m$^{-1}$ K$^{-1}$ at 300 K, thus matching well with our experimental results. As also seen in our calculations, the low thermal conductivity of $SnS_2$ can be attributed to low group velocity, a shorter lifetime of the phonons, and strong anharmonicity in the crystal.

**Techniques and methods**

*Experimental details:*

Flakes of desired thickness are prepared by mechanical exfoliation method from highly pure single crystals of 2H-SnS$_2$ (Manchester nanomaterials) using scotch tape and transferring onto Si substrates coated with ~ 300 nm thick SiO$_2$ film (SiO$_2$/Si). Samples were prepared on a substrate pre-cleaned in acetone, isopropyl alcohol (IPA), and deionized water. The layer numbers of the flakes are verified using Raman spectroscopy as well as atomic force microscopy (AFM). The monolayer is found to be ~ 0.7 nm thick from the height profile of the topographic image of AFM (Agilent 5000) in contact mode. Raman measurements were performed in the back-scattering geometry by using a LABRAM HR Evolution Raman spectrometer (1800 rulings/mm), fitted with a microscope having 50× (NA = 0.5), 20× (NA = 0.4), and 100× (NA = 0.6) objectives, and coupled to an air-cooled charge-coupled device (CCD) detector. All the measurements were carried out with 532 nm laser excitation. The pixel resolution was ~ 0.37 cm$^{-1}$. To measure the laser spot size for different objectives, we have performed a Raman linear mapping across a sharp-edged Au strip deposited on a SiO$_2$/Si substrate. The spot radii were determined to be 0.64 μm (for a 50× objective lens), 0.94 μm (for a 20× objective lens), and 0.34 μm (for a 100× objective lens) using a technique [26,34] discussed in the Supplemental note S1 of the Supplemental material [35]. The temperature-dependent Raman measurements were performed using a fixed laser power (~ 1 mW) inside a LINKAM heating stage (Model HFS600E-PB4) and varying the temperature from 300 to 420 K in steps of 10 K. The laser power-dependent measurements were performed at room temperature, varying the laser power upto an upper limit (~ 6 mW for 50× and 20× lenses and ~ 1.5 mW for 100× objective) to avoid any laser-induced degradation of the 2D layers.

*Computational details:*

Density functional theory (DFT) [36] based calculations were carried out using Quantum Espresso (QE) code [37]. The generalized gradient approximation (GGA) within the Perdew-Burke-Ernzerhof (PBE) [38] flavor was used for the electronic exchange and correlation functional. A 24×24×1 dense k-point sampling in the hexagonal Brillouin zone (BZ) was used for the geometry relaxation of the unit cell, whereas the cut-off energy for the plane wave expansion was kept at 60 Ry. Geometry relaxations were performed until the maximum force in the individual atom was reduced to a value of 0.01 eV/Å. The effect of the monolayer was implemented in our structure by creating a vacuum of length 16 Å along the z-direction to avoid interlayer interactions. The lattice thermal conductivity was calculated in PHONO3PY [39] interfaced with the QE package. To calculate the third-order anharmonic and second-order harmonic interatomic force constants (IFCs), a 2×2×1 supercell was used with 9×9×1 k point sampling. The interaction cutoff used in the calculation of third-order IFCs in PHONO3PY was 5.29 Å. For calculating lattice thermal conductivity ($\kappa$) 72×72×1 k point sampling was used. The detailed methodology of the calculation can be found in our previous work [40]. For 2D materials, the calculated value of lattice thermal conductivity ($\kappa$) depends on the length of the unit cell along the z-direction. So, the $\kappa$ value obtained from PHONO3PY has been normalized by multiplying a factor $L_z/d$, where $L_z$ is the total length of the unit cell along the z-direction and d is the thickness of the materials. In this paper, $L_z$ = 20 Å and d = 6.98 Å for monolayer are obtained by considering the thickness of one single layer in the bulk unit cell of SnS$_2$ that matches well with the monolayer thickness of ~ 0.7 nm obtained from the AFM measurements.

**Results and Discussions**

The bulk SnS$_2$ belongs to the space group $P\bar{3}m1$ (No. 164) with a hexagonal unit cell, as shown in FIG. S1 of the Supplemental material [35], with lattice parameters a = b = 3.72 Å [23,41,42]. FIG. 1(a) shows the optical images of the experimentally investigated flakes of different thicknesses. The thicknesses of the flakes were confirmed by AFM measurements. FIG. 1(b) shows the topographic images and the corresponding height profiles for the flakes. As can be seen from the height profiles, the thickness of 1L SnS$_2$ is ~ 0.7 nm, while those of 2L and 5L SnS$_2$ are ~ 1.4 and ~ 3.2 nm, respectively (see Supplemental note S3 and FIG S3 in Supplemental material for more details) [35]. FIG. 1(c) shows the room temperature Raman spectra of the flakes of 2H SnS$_2$ with varying layer thickness. At the center (Γ - point) of the Brillouin zone of 2H-SnS$_2$, there are nine normal modes of vibration, out of which three are acoustic and six are optical, which may be represented using the following irreducible representation: $\Gamma_{bulk} = A_{1g} + E_g + 2A_{2u} + 2E_u$, where infrared-active and acoustic modes are represented by $A_{2u}$ and $E_u$ symmetry while the $A_{1g}$ and $E_g$ modes are Raman active. [18,43] It may be noted that the 2H-SnS$_2$ is structurally equivalent to the 1T polytype of MoS$_2$ having the point group $D_{3d}$. Similar to 1T-MoS$_2$ the space group in the N layer SnS$_2$ (N = even or odd) is the same as the bulk 2H-SnS$_2$. Therefore, the N layer (even or odd) is represented by the same irreducible representation as bulk 2H-SnS$_2$. The $A_{1g}$ mode (out-of-plane vibration of the S atom) appears at 314 cm$^{-1}$ whereas the $E_g$ mode (in-plane vibrations of Sn and S atoms) could not be detected, possibly due to a low scattering cross section. The intensity of the $A_{1g}$ mode increases with the increase in the number of layers [44].

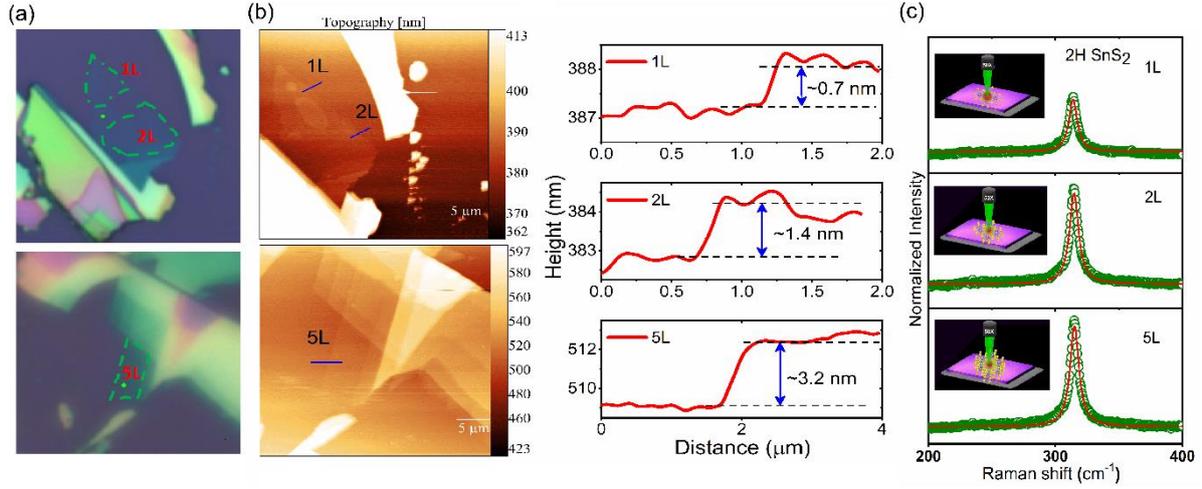

**FIG. 1.** (Color online) (a) Optical microscopic images of 1L, 2L, and 5L 2H-SnS$_2$. (b) AFM topographic image with the corresponding height profile for the 1L, 2L, and 5L flakes. (c) Room temperature Raman spectra for 1L, 2L, and 5L SnS$_2$. The inset figures show the schematic diagrams depicting the backscattering geometry for the respective layers.

FIG. 2 shows contour (color) maps where we can observe a redshift of the $A_{1g}$ mode in all the flakes with the increase in temperature and the laser power (corresponding Raman spectra are shown in FIG. S4 in Supplemental material [35]). An increase in the laser power at room temperature increases the local temperature of the layers. The observed redshift of the $A_{1g}$ mode with both temperature and laser power is due to the thermally induced softening of the bonds [45]. FIG. 3 shows the shift in frequency of the $A_{1g}$ mode with temperature and laser power. In the measured temperature and laser-power range, the $A_{1g}$ phonon mode can be fitted with a linear equation $\omega(T) = \omega_0 + \chi_T T$ and $\omega(P) = \omega_{p_0} + \chi_p P$ where $\omega_0$ is the mode frequency at 0 K, $\chi_T$ is the first-order temperature coefficient, T is the absolute temperature, $\omega_{p_0}$ is the frequency at room temperature without any laser-induced heating, $\chi_p$ is the first-order power coefficient, and P is the laser power. The first-order temperature and power coefficients, $\chi_T = \left(\frac{d\omega}{dT}\right)_P$ and $\chi_P = \left(\frac{d\omega}{dP}\right)_T$, respectively, for the Raman active $A_{1g}$ mode in the various investigated flakes are listed in Table 1. To measure the in-plane thermal conductivity ($\kappa$) of SnS$_2$ and the interfacial thermal conductance per unit area (g) between the SnS$_2$ layer and the SiO$_2$/Si substrate, we have used the following heat diffusion equation in the cylindrical coordinate system, assuming that no heat loss occurs through the surrounding atmosphere [26,31,32,35,46].

$$\frac{1}{r}\frac{d}{dr}\left(r\frac{dT}{dr}\right) - \frac{g}{\kappa l}(T - T_a) + \frac{q}{\kappa} = 0 \qquad (1)$$

where, $T$ is the temperature at position $r$ on the layer, $T_a$ is the ambient temperature, $l$ is the thickness of the layers, and $q$ represents the volumetric heat coefficient.

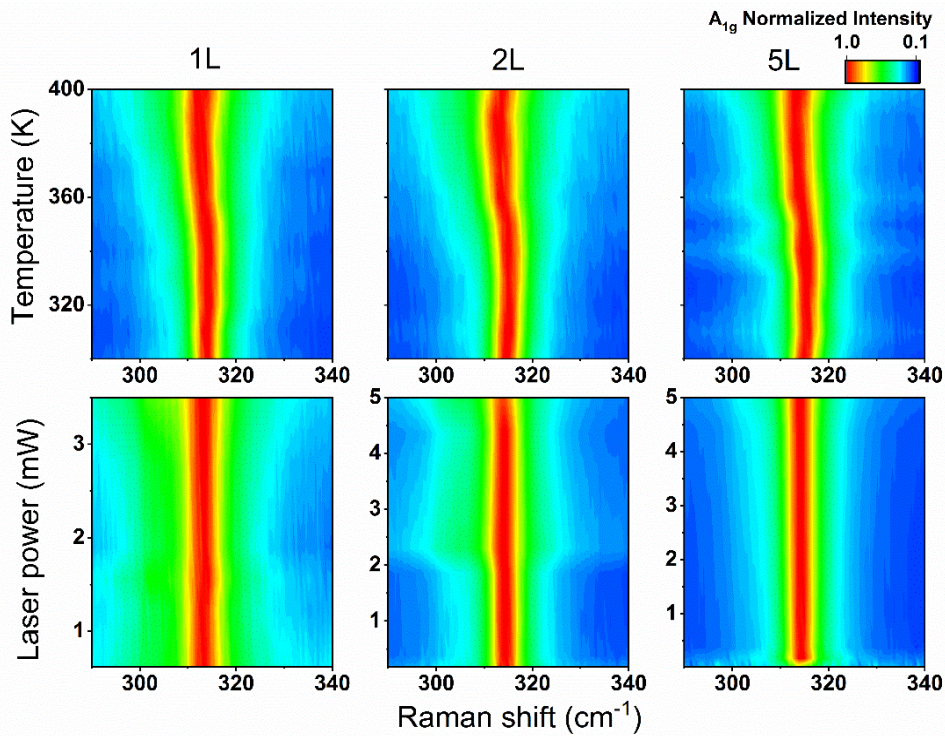

**FIG 2:** (Color online) Contour (color) maps showing the intensity profiles of the $A_{1g}$ mode and its frequency dependence for the 1L, 2L, and 5L samples with the temperature variation (top panels) and the laser-power variation (bottom panels).

The above equation can be used to estimate the interface thermal resistance $R_m$ (see Supplemental note S5 of the Supplemental material [35]) which is related to the measured $\chi_T$ and $\chi_p$ as:

$$R_m = \frac{\partial \theta_m}{\partial Q} = \frac{\partial \omega}{\partial Q}\frac{\partial \theta_m}{\partial \omega} = \chi_P (\chi_T)^{-1} \quad (2)$$

where, $\theta_m$ and $Q$ represent the measured sample temperature and the total absorbed laser power. The κ and g values for the different flakes are obtained by using equation-2 (see Supplemental note S5 of the Supplemental material for further details [26,35]). In order to correctly estimate the thermal properties, it is essential to measure the laser spot radius, the thickness of the investigated flakes, and the absorption coefficients [47,48] accurately, which are discussed in the Supplemental note S6 of the Supplemental material [35].

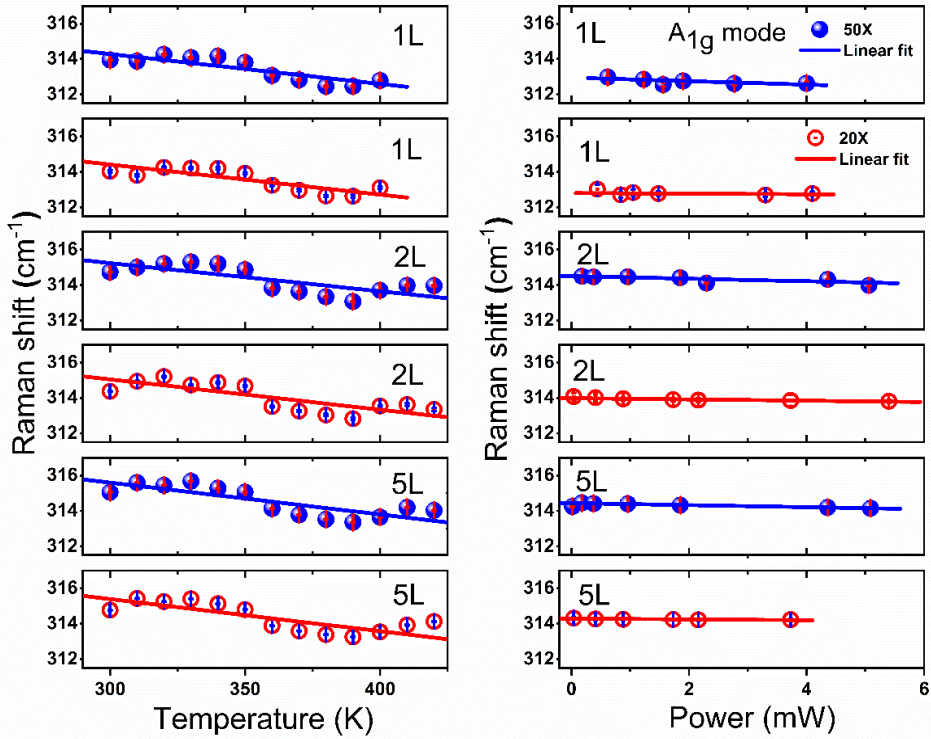

**FIG. 3.** (Color online) Temperature and power dependence of $A_{1g}$ Raman modes for 1L, 2L, and 5L SnS$_2$ flakes. The slopes give the first-order temperature and power coefficients which are enlisted in Table 1. Filled circles (blue) represent the data recorded using 50× objective while the open circles (red) correspond to the data obtained using 20× objective.

*Table1: First-order temperature and power coefficients of the various investigated flakes.*

| Sample | Modes | Objective | $\chi_T$ (cm$^{-1}$/K) | $\chi_P$ (cm$^{-1}$/mW) |
|---|---|---|---|---|
| **1L** | $A_{1g}$ | 50× | -0.019 | -0.090 |
|  |  | 20× | -0.016 | -0.039 |
| **2L** | $A_{1g}$ | 50× | -0.015 | -0.084 |
|  |  | 20× | -0.017 | -0.045 |
| **5L** | $A_{1g}$ | 50× | -0.018 | -0.045 |
|  |  | 20× | -0.016 | -0.020 |

FIG. 4 represents the simultaneously measured κ and g values of 2H-SnS$_2$ with varying thickness. The in-plane thermal conductivity of the 1L flake is κ = 3.20 ± 0.57 W m$^{-1}$ K$^{-1}$, which gradually increases with the thickness of the flake to about 5.00 ± 0.90 W m$^{-1}$ K$^{-1}$ for the 5L flake. Similarly, the cross-plane interface thermal conductance per unit area (g) increases with layer thickness from 0.53 ± 0.09 MW m$^{-2}$ K$^{-1}$ for the monolayer flake to 3.60 ± 0.5 MW m$^{-2}$ K$^{-1}$ for the 5L SnS$_2$ flake. The errors in the estimation of the thermal conductivity were calculated by the root-sum-square-error propagation approach. (see Supplemental note S8 of the Supplemental material for more details [35]). Importantly, these estimates (of κ and g) are insensitive to (independent of) the laser spot size used during the measurements. This is because the lateral heating length ($L_h \sim 65\ nm$), which is the characteristic length up to which the heat flows before it sinks, is much shorter than the spot size (~ 300 – 900 nm in our case) [49] (see supplemental note S7 [35] for more details). For substrate-supported 2D layers, the heat flows predominantly in two directions (neglecting the convection and radiation losses): one along the radially outward direction within the plane from the center of the laser spot (which is defined by κ) and the other in the cross-plane (out-of-plane) direction toward the SiO$_2$/Si substrate (defined by g). The κ and g values, thus measured, are very low as compared to other 2D layered materials like graphene (600 W m$^{-1}$ K$^{-1}$) [26,27], and h-BN (280 W m$^{-1}$ K$^{-1}$) [26] as well as TMDCs like MoS$_2$ (62 W m$^{-1}$ K$^{-1}$) [31] MoSe$_2$ (59 W m$^{-1}$ K$^{-1}$) [32], and WS$_2$ (32 W m$^{-1}$ K$^{-1}$) [33].

The very low thermal conductivity of monolayer and thin SnS$_2$ can be attributed to the suppression of the flexural phonons because of the non-planar structure as compared to planar graphene. As can be seen in fig. 4, though the thermal conductivity does not show a dramatic change with the layer thickness, it exhibits a significant and systematic increase as the thickness increases from monolayer to five layers (i.e 5L). When we increase the layer number to 2L or 5L, only the bottom layer at the interface couples to the substrate while the upper layer(s) of SnS$_2$ are weakly (or not) coupled to the substrate. This leads to an improvement in the lifetime of the in-plane phonons of SnS$_2$ (as compared to that in monolayer); as a result, the in-plane thermal conductivity increases with increasing layer thickness [26,50]. The in-plane thermal conductivity is also influenced by the phonon boundary scattering. As the layer number increases the phonon boundary scattering decreases which in turn improves the in-plane thermal conductivity [51–54]. As a result, we can see a significant improvement in the in-plane thermal conductivity with increasing layer number (thickness) as compared to that in the monolayer SnS$_2$. On the other hand, the interface thermal conductance per unit area also increases with increasing thickness due to an effective reduction of the surface roughness seen by the top layer(s) of SnS$_2$ supported on SiO$_2$/Si thus increasing the effective surface area of contact for out-of-plane heat dissipation [26,55,56]. The effective contact area between the 2D layers and substrate plays a crucial role in improving the interfacial thermal conductance per unit area (g) [26,57]. With the increase in the layer number, the contact area between the layer and the substrate increases and, hence, the layers become flatter, which, as a result, improves the out-of-plane heat dissipation channel and increases the g [26,55,57–59]. Further understanding of the origin of the very low thermal conductivity can be obtained from our DFT calculations, as discussed below.

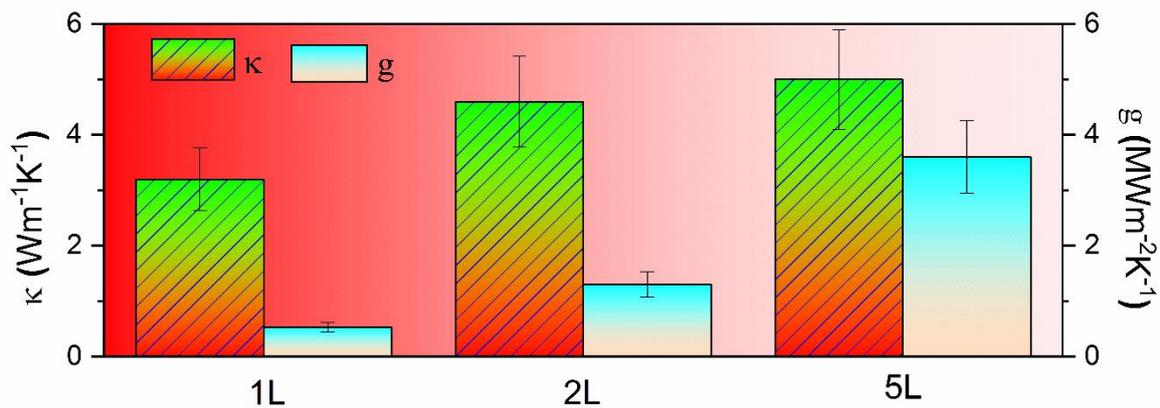

**FIG. 4.** (Color online) Bar diagram comparing the in-plane thermal conductivity ($\kappa$) (left axis) and the interface thermal conductance per unit area (g) (right axis) values for the three different thicknesses of 2H-SnS$_2$.

**An insight from density functional theory:**

The phonon band structure for monolayer SnS$_2$ has been calculated using density functional theory, as shown in FIG. 5(a). There are three acoustic branches (in-plane LA and TA and out-of-plane ZA) and six optical branches in the phonon dispersion curves. Among the optical branches, there are two Raman active modes, $E_g$ (at ~185 cm$^{-1}$) (due to the in-plane vibrations), and $A_{1g}$ (due to out-of-plane symmetric vibration). The out-of-plane $A_{1g}$ mode appears at 301 cm$^{-1}$ in our calculations which is experimentally observed at 314 cm$^{-1}$ in the Raman spectra as discussed above. The $A_{2u}$ mode at ~ 336 cm$^{-1}$ (due to the out-of-plane asymmetric vibration) and the $E_u$ mode at ~ 192 cm$^{-1}$ (due to in-plane vibrations) are IR-active phonons. The acoustic modes arise due to the vibration of Sn atoms, while the optical branches originate from the vibration of S atoms, as can be seen from the phonon density of states in FIG. 5(a). It is to be noted that there is a very small phonon band gap of ~ 8 cm$^{-1}$ between the acoustic and optical branches as can be seen in the phonon dispersion of 1L-SnS$_2$ which is in sharp contrast with a large gap between the acoustic and optical branches as reported for other TMDCs such as 1L-MoS$_2$ (~ 49 cm$^{-1}$) [60,61] and 1L-WS$_2$ (~ 110 cm$^{-1}$) [40,62]. The calculated lattice thermal conductivity ($\kappa$) of 1L SnS$_2$ as a function of temperature has been shown in FIG. 5(b).

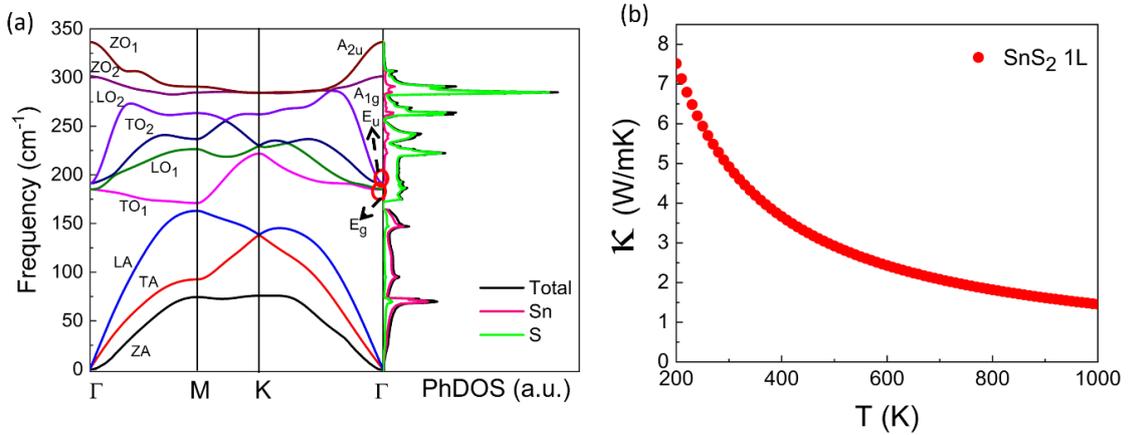

**FIG. 5.** (Color online) (a) Phonon dispersion curves and phonon density of states (PhDOS) of monolayer SnS$_2$. (b) Lattice thermal conductivity with respect to the temperature for the monolayer SnS$_2$.

Our calculated value of $\kappa$ is 4.92 W m$^{-1}$ K$^{-1}$ at 300 K, which is in good agreement with our experimental value of 3.20 W m$^{-1}$ K$^{-1}$. It may be noted that the value of $\kappa$ of SnS$_2$ obtained in our calculation and experiment is comparable to that in some of the TMDCs such as 1L-HfS$_2$ [63] and 1L-ZrS$_2$ [64] but much lower than that in 1L-MoS$_2$ [65] and 1L-WS$_2$ [40]. An understanding of the origin of the very-low value of thermal conductivity in SnS$_2$ can be developed based on its phonon band structure, phonon lifetime, and group velocity. The lattice thermal conductivity can be expressed as [66]:

$$\kappa^i = \sum_{q,j} \kappa^i_{q,j} = \sum_{q,j} C_v(w_{q,j}, T) \Lambda^i(w_{q,j}, T) \upsilon^i(w_{q,j}, T) \qquad (3)$$

Where, $C_v(w_{q,j}, T)$, $\Lambda^i(w_{q,j}, T)$, and $\upsilon^i(w_{q,j}, T)$ represent the mode-dependent specific heat, phonon mean free path (MFP), and group velocity, respectively.

An explanation for the low value of thermal conductivity in 1L-SnS$_2$ can be found from the phonon dispersion curves. Due to a very small phonon gap (~ 8 cm$^{-1}$) in SnS$_2$, as shown in FIG. 5(a), it is possible that the acoustic LA branch couples with the low-frequency optical mode (TO$_1$). A coupling between the modes would lead to an increased phonon scattering, resulting in a reduced group velocity. According to the anharmonicity theory, the optical phonon decays into acoustic phonons, giving rise to a finite lifetime for the phonons [67]. The coupling between the optical and acoustic branches (due to the small band-gap) can therefore, ensure stronger anharmonic effects. The mode-dependent group velocity as a function of phonon frequency (in THz) has been shown in FIG. 6(a). It is important to note that the group velocity of the acoustic branch of SnS$_2$ is about one order less than the group velocity of acoustic modes in graphene [68].

Further, the group velocity of optical branches is lower than that of acoustic branches. Therefore, as can be seen in FIG. 6(b), the acoustic modes contribute almost 95% of the total thermal conductivity, whereas the contribution of the optical modes is about 5%. Among the acoustic branches, the out-of-plane ZA and in-plane LA modes contribute the most. We have further calculated the heat capacity as a function of temperature (see FIG. S7 in Supplemental note S9 of the Supplemental material [35]), which is found to be low. A low value of heat capacity (~ 32 J K$^{-1}$ mol$^{-1}$ at 300 K) can be associated with a low phonon density of state. Therefore, the low values of group velocity and heat capacity would result in a low thermal conductivity in SnS$_2$.

The Grüneisen parameter is one of the important factors to determine the lattice thermal conductivity as it represents the degree of anharmonicity in the crystal. The mode-dependent

Grüneisen parameter as a function of frequency has been shown in FIG. 6(c). The Grüneisen parameter ranges between a large negative value and a positive value for the ZA branch as a function of frequency and this phenomenon has also been reportedly seen in other 2D materials [69–71]. It is also observed that the Grüneisen parameters of the acoustic branches are much greater than unity and much higher than those of the optical branches. The large absolute values of the Grüneisen parameters in 1L-SnS$_2$ are clear indications of the presence of a very strong anharmonicity in the crystal that can result in the observed low value of lattice thermal conductivity.

The calculated phonon lifetime ($\tau$) of the normal modes of 1L SnS$_2$ has been shown as a function of frequency in FIG. 6(d), where the color bar represents the phonon density. It is important to note that the calculated highest value of the lifetime of phonons is around 3 ps in 1L-SnS$_2$, which is orders of magnitude lower than that in MoS$_2$ and WS$_2$ [62], thus suggesting the presence of strong phonon-phonon scattering in SnS$_2$. The phonon lifetime is associated with the mode-dependent mean free path as [66]:

$$\Lambda_{q,j}^i = v_{q,j}^i \tau \qquad (4)$$

As discussed above, the mode coupling (between LA and TO$_1$) results in low group velocity ($v_{q,j}^i$) (FIG. 6(a)) and short lifetime ($\tau$) (see FIG. 6(d)) of the phonons in 1L-SnS$_2$, thus reducing the phonon mean free path ($\Lambda_{q,j}^i$). Therefore, the low values of the phonon mean free path, group velocity, as well as heat capacity give rise to a very low thermal conductivity in SnS$_2$.

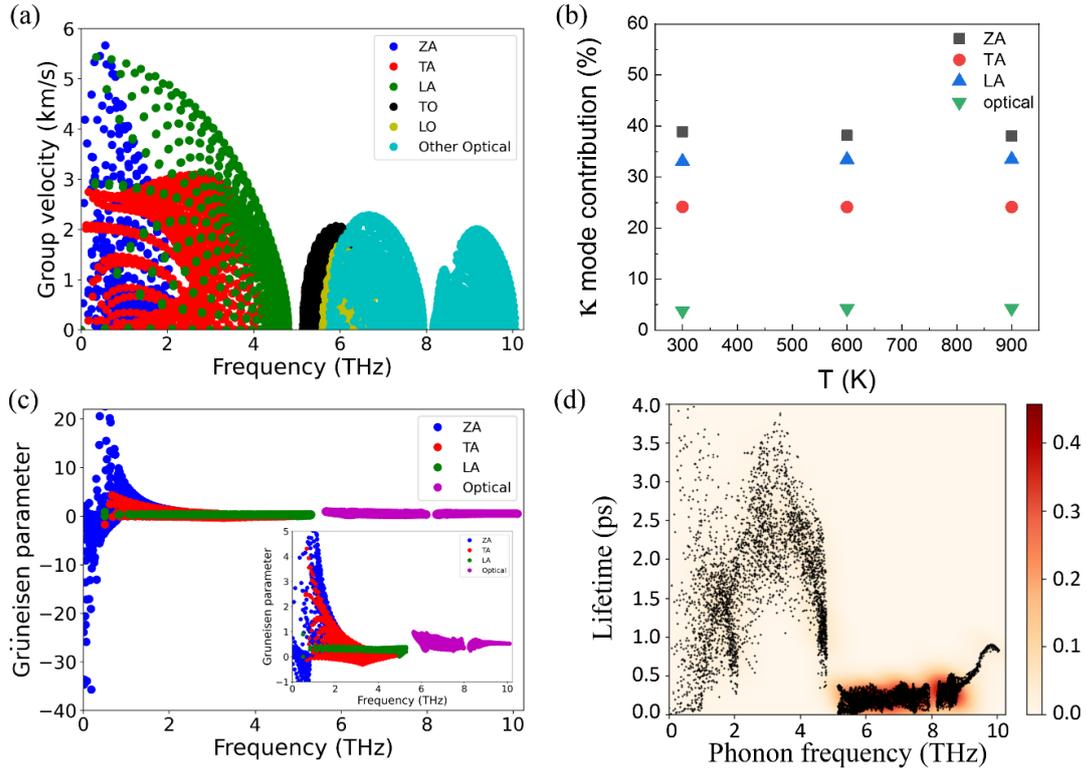

**FIG. 6.** (Color online) (a) Phonon group velocity for all the modes of monolayer $SnS_2$. (b) Percentage contribution of acoustic and optical modes in lattice thermal conductivity with respect to temperature for the monolayer $SnS_2$. (c) Mode-dependent Grüneisen parameters for the monolayer $SnS_2$ and the inset shows the magnified view of the region from -1 to 5 for clarity. (d) Lifetime of the phonon modes for monolayer $SnS_2$ and color bar represents the density of phonon modes in the frequency-lifetime plot.

As discussed above, the low mean free path contributes to the low value of thermal conductivity in $SnS_2$. Therefore, it is important to investigate the lateral size effect on the thermal properties to get a better insight. The (lateral) size effect of the 2D layer can be investigated by studying the cumulative lattice thermal conductivity (CLTC) as a function of phonon mean free path. The fig. 7 shows the variation of the calculated normalized CLTC as a function of phonon MFP at 300 K. The CLTC increases as the MFP increases and becomes saturated at a MFP of ~ 588 nm where CLTC reaches almost 98% of its highest value. To get a better insight we have fitted the CLTC data to a single parametric function [72,73], $\kappa(\Lambda < \Lambda_{max}) = \frac{\kappa_{max}}{1+\Lambda_0/\Lambda_{max}}$, where $\kappa_{max}$ and $\Lambda_{max}$ are the maximum cumulative lattice thermal conductivity and corresponding maximum MFP, respectively, and $\Lambda_0$ is the parameter to be determined from the fitted curve. When the sample size is equal to or greater than $\Lambda_{max}$, the cumulative lattice thermal conductivity saturates. By fitting the CLTC data, we find the value of $\Lambda_0 \sim 32$ nm for monolayer $SnS_2$ that matches well with previous reports [11,23,74]. As can be seen, 90% of the total cumulative lattice thermal conductivity is contributed by the phonons with MFP less than 153 nm, which means that if the sample size is greater than 153 nm then there will be a negligible effect of sample size on the lattice thermal conductivity. However, the lattice thermal conductivity can be reduced by lowering the lateral dimension to below ~ 153 nm by appropriate nano structuring that may have potential application in nano devices.

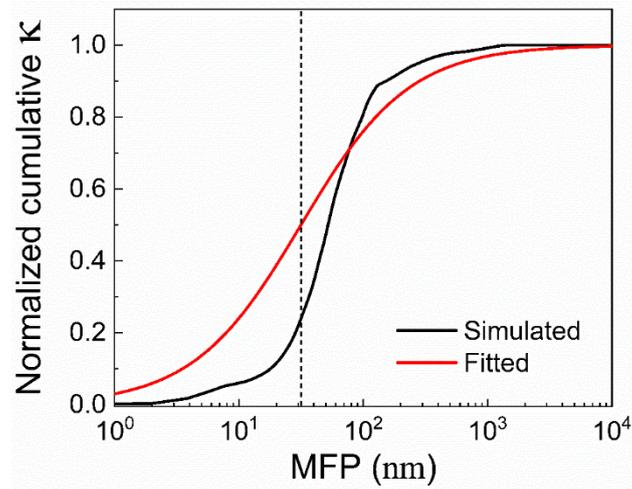

**FIG. 7.** (Color online) Variation of normalized cumulative lattice thermal conductivity as a function of phonon mean free path.

*Table 2. Summary of In-Plane Thermal Conductivity and interfacial thermal conductance of Similar Layered Materials.*

| Materials | In plane thermal conductivity κ (W m$^{-1}$ K$^{-1}$) | Interfacial thermal conductance per-unit area g (MW m$^{-2}$ K$^{-1}$) | remarks | References |
|---|---|---|---|---|
| SnSe$_2$ | 1.80 - 2.90 | - | 5 - 15 Layer polycrystalline SnSe$_2$ nanofilms | [52] |
| SnS$_2$ | 3.20 | 0.53 | Monolayer (Opto thermal Raman) | This work |
| SnS$_2$ | 3.45 | - | 16nm thick (experiment) | [21] |
| SnS$_2$ | 4.60 | 1.30 | 2 Layer (Opto thermal Raman) | This work |
| SnS$_2$ | 4.92 | - | Monolayer (Theory) | This work |
| SnS$_2$ | 5.00 | 3.60 | 5 Layer (Opto thermal Raman) | This work |
| SnS$_2$ | 6.41 | - | Theory | [23] |
| SnS$_2$ | 8.20 | - | Theory | [11] |
| SnS$_2$ | 10 | - | Bulk (experiment) | [24] |
| WS$_2$ | 32 | - | 1L exp | [33] |
| MoSe$_2$ | 59 | 0.09 | 1L exp | [32] |
| MoS$_2$ | 62 | 1.94 | 1L exp | [31] |

The lattice thermal conductivity heavily depends on some of the important factors like Debye temperature, phonon group velocity, and Grüneisen parameter. Table-2 summarizes the thermal properties of SnS$_2$ comparing our result and available reports. We have shown that the group velocity of the acoustic phonons of monolayer SnS$_2$ is one order less than that in graphene [23,45,75,76]. Further, the acoustic phonons are the main contributors for the heat dissipation [75]. A large value of the Grüneisen parameter of the acoustic modes implies a strong phonon anharmonicity. Shafique *et al.* [23] showed that the calculated Debye temperature of SnS$_2$ is smaller than that in the other two-dimensional materials like graphene, h-BN, and MoS$_2$. A low value of Debye temperature implies that at room temperature a greater number of phonon modes are activated, thus causing an increment in the phonon scattering rate, which in turn decreases the lattice thermal conductivity [23]. We find multiple compelling reasons, such as low group velocity, shorter phonon lifetime, large Grüneisen parameter, and low heat capacity, that lead to the very low value of thermal conductivity in $SnS_2$. Lee *et al.* [21] reported in their work that with the decreasing thickness the electrical conductivity increases, while the thermal conductivity decreases because of the surface phonon scattering, thus making 2H-$SnS_2$ a promising candidate as a thermoelectric material [21]. Hence, our experimental and theoretical findings provide further evidences justifying the origin of the very low thermal conductivity in $SnS_2$, thus corroborating previous reports [11,21,22].

**Conclusion:**

We have reported the in-plane thermal conductivity (κ) as well as the out-of-plane interface thermal conductance per unit area (g) between the layer and substrate ($SiO_2$/Si) measured by optothermal Raman technique for three different thicknesses of 2H-$SnS_2$. The in-plane thermal conductivity is found to be ~ 3.20 W m$^{-1}$ K$^{-1}$, while the out-of-plane thermal conductance per unit area is ~ 0.53 MW m$^{-2}$ K$^{-1}$ for monolayer $SnS_2$ and shows an appreciable increase with increasing thickness. We have also calculated the lattice thermal conductivity using DFT for the monolayer $SnS_2$ that matches well with our experimental results. The very low value of lattice thermal conductivity of $SnS_2$ arises due to low group velocity, shorter phonon lifetime, low heat capacity, and strong anharmonicity in the crystal. Based on our results, we believe that thin layers of $SnS_2$ can be a good candidate to explore its thermoelectric device applications because of its very low lattice thermal conductivity, which is an important requirement to achieve a high ZT coefficient.

**Acknowledgments:**

We sincerely acknowledge the funding from DST/SERB (Grant No. ECR/2016/001376 and Grant No. CRG/2019/002668), Nanomission (Grant No. SR/NM/NS-84/2016(C)), MHRD (STARS/ APR 2019/ PS/ 662/FS) and DST-FIST (Grant No. SR/FST/PSI-195/2014(C)). The authors acknowledge the Central Instruments Facility, IISER Bhopal, for AFM facility. We are thankful to IIT Jodhpur for providing the necessary computational facility and the Ministry of Human Resource and Development (MHRD) for the infrastructure.

**Supplemental material**

**Low Thermal Conductivity and Interface Thermal Conductance in SnS$_2$**

Saheb Karak[1], Jayanta Bera[2], Suvodeep Paul[1], Satyajit Sahu[2], and Surajit Saha[1*]

[1]Department of Physics, Indian Institute of Science Education and Research, Bhopal, 462066, India

[2]Department of Physics, Indian Institute of Technology Jodhpur, Jodhpur 342037, India

*Correspondence: surajit@iiserb.ac.in

This Supplemental material contains additional discussions on laser spot radius measurements for 50×, 20×, and 100× objective lenses, the crystal structure of monolayer 2H-SnS$_2$, thickness measurement using atomic force microscopy (AFM), temperature and power-dependent Raman spectra, optothermal method for measurement of κ and g, absorption coefficient measurements of 2H-SnS$_2$ thin layers using AFM and Raman, dependence of thermal conductivity (κ) and interfacial thermal conductance per unit area (g) on laser spot sizes, Error calculations, and heat capacity calculation using density functional theory (DFT).

**Supplemental Note S1: Determination of spot radius for 50×, 20×, and 100× objective**

We have determined the laser spot radius for 50×, 20×, and 100× objectives using a procedure similar to a knife-edge technique [1,2]. A sharp-edged thin strip of gold was deposited on the SiO$_2$/Si substrate and then a linear Raman mapping was done across the sharp edge of the gold stripe in the range of 450 cm$^{-1}$ to 550 cm$^{-1}$. FIG. S1 shows the corresponding intensity profile of the Si peak along with the linear map. The intensity profile was fitted with the function:

$$I(x) = \frac{I_0}{2}\left(1 + erf\left(\frac{x - x_0}{w_0}\right)\right)$$

The fitted intensity profile provides an estimate of the spot radius ($r_o$) for 50× ($r_o \sim 0.64$ μm), 20× ($r_o \sim 0.94$ μm), and 100× ($r_o \sim 0.34$ μm) objective lenses.

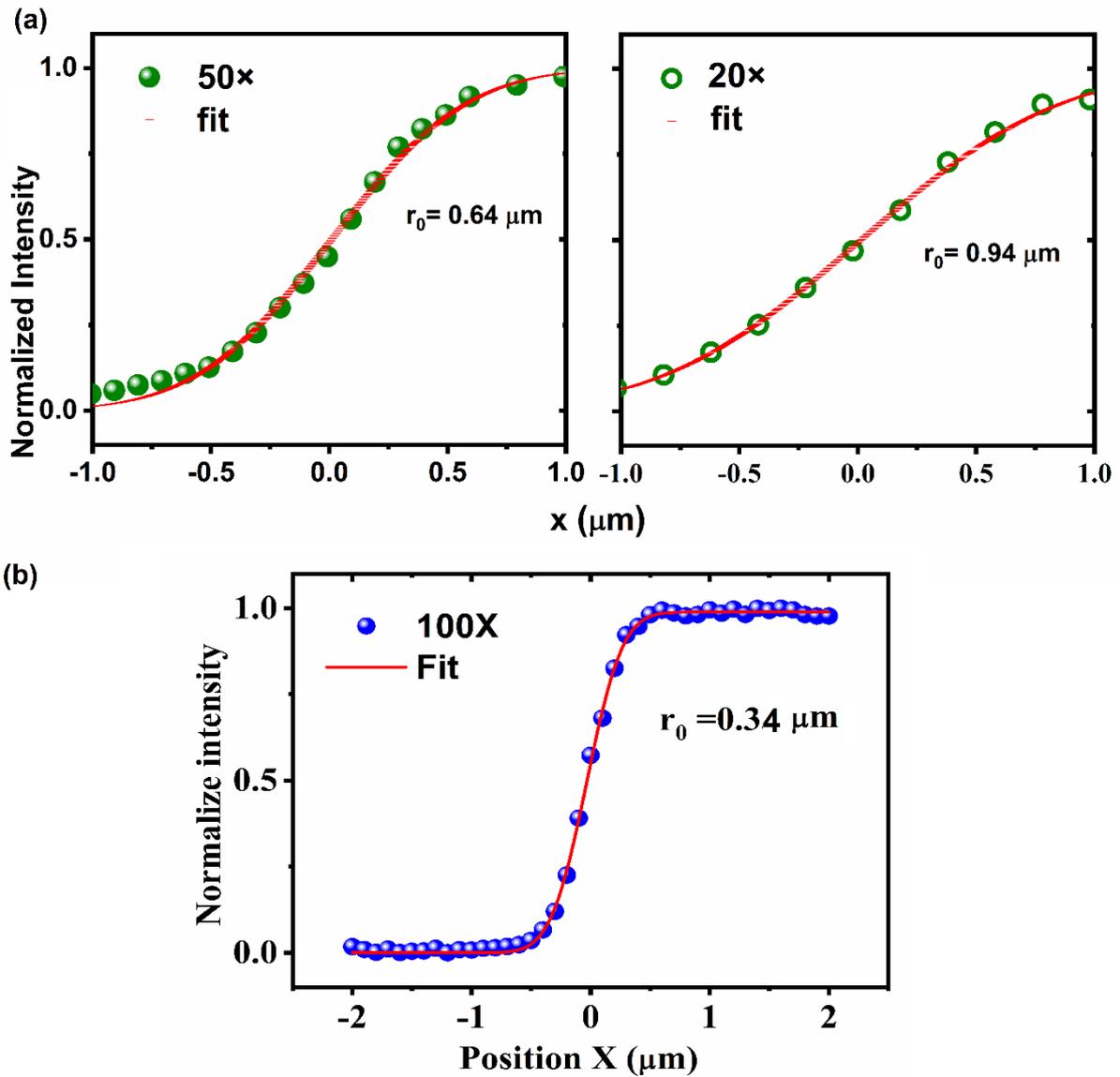

**FIG. S1.** (a) The intensity profile of the Si mode across the sharp gold strip deposited on the SiO$_2$/Si substrate in order to estimate the spot radii corresponding to the 50× (left) and 20× (right) objective lenses, and (b) for the 100× objective lenses.

**Supplemental Note S2: Crystal structure of SnS$_2$**

Depending upon the stacking sequence of the individual layers SnS$_2$ can be categorized into seventy polytypes [3].

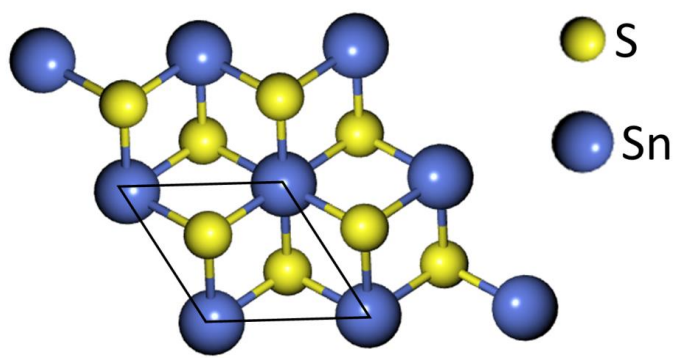

**FIG. S2.** Top view of the crystal structure of monolayer 2H-SnS$_2$.

The stable structure is 2H-SnS$_2$, where Sn is sandwiched between two Sulphur layers. The structure of 2H-SnS$_2$ is very similar to that of 1T-MoS$_2$. To form a bulk SnS$_2$ crystal, monolayers of 2H-SnS$_2$ stack exactly on top of other. FIG. S2 gives the crystal structure of monolayer SnS$_2$.

**Supplemental Note S3: Thickness measurement**

The layer thickness of $SnS_2$ flakes have been measured using AFM. FIG. S3 (a) shows the optical image of the 6L and 7L (L = layers) samples. FIG. S3(b) shows the topographic image for the 6 and 7 layers while FIG. S3 (c, d) give the height profile for the same. The thickness of 6 L $SnS_2$ is ~ 3.85 nm and of 7L flake is ~ 4.50 nm.

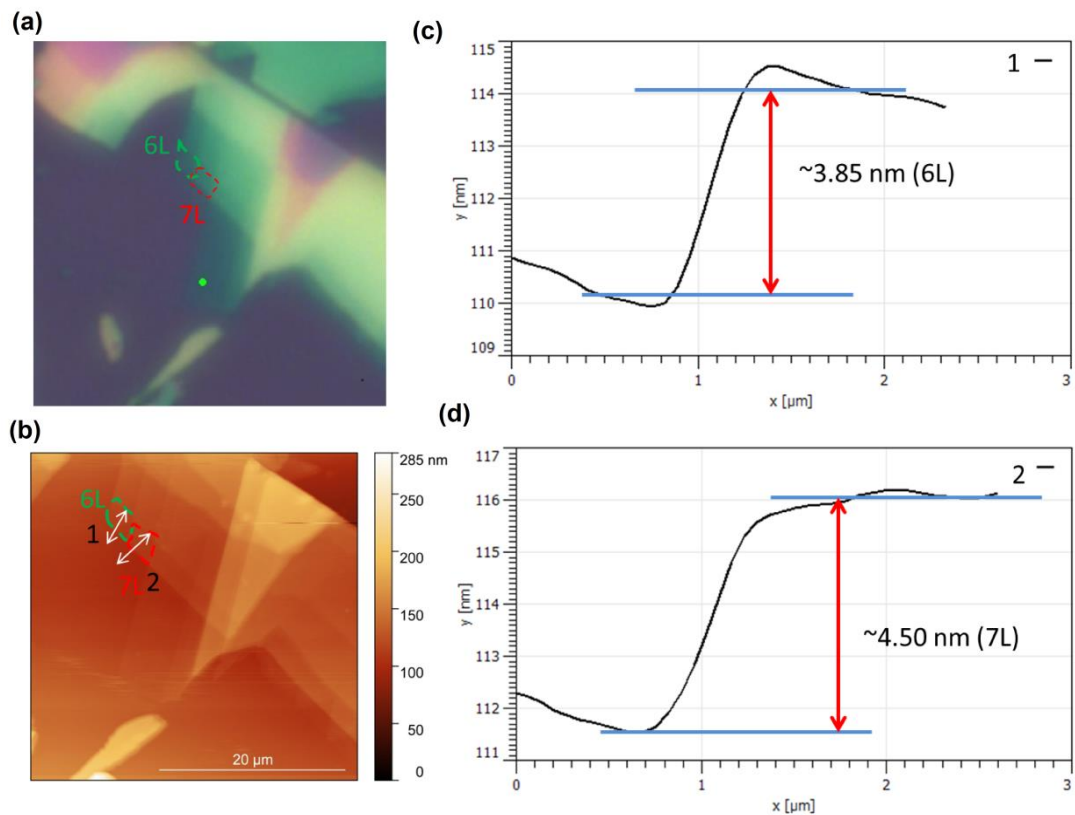

**FIG. S3.** (a) Optical image for the 6L and 7L flakes. (b) AFM topographic image for the same layers. (c, d) height profiles for the 6L and the 7L flakes, respectively.

**Supplemental Note S4: Temperature and power-dependent Raman spectra**

Temperature and power-dependent Raman studies were recorded for all the flakes. FIG. S4 (a and b) show the temperature and power-dependent spectra of the respective flakes. It was observed that all modes show a red-shift with an increase in temperature as well as power. The red-shift can be attributed to phonon anharmonicity [4]. The power-dependent spectra also show a similar red-shift because the laser-power induces a local heating of the sample, thereby increasing the temperature near the exposed part of the sample. We have performed laser power-dependent experiments in a range from 0.03 mW to 6.00 mW. On the other hand, the temperature-dependent measurements are performed at a fixed low laser power (~ 1 mW) to minimize laser-induced heating. Hence, the signal-to-noise ratio in the temperature-dependent spectra remains almost comparable.

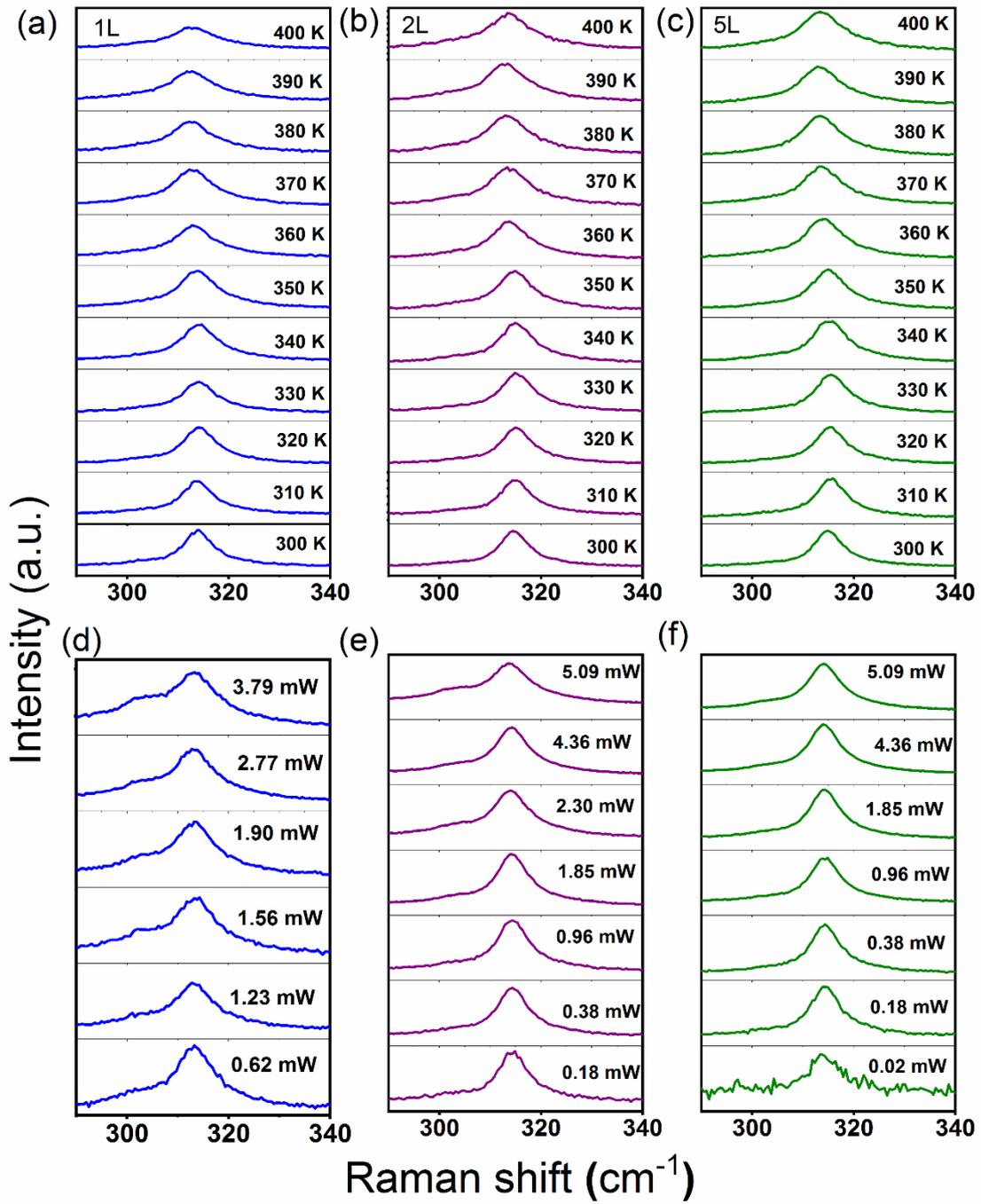

**FIG S4:** Comparison of temperature-dependent and power-dependent Raman spectra for 50× lens, showing the $A_{1g}$ mode of $SnS_2$ (a), (d) for 1L, (b), (e) for 2L and (c), (f) for 5L. The mode shows a red-shift with the increase in temperature as well as laser power.

**Supplemental Note S5: The Optothermal method and analysis details**

The optothermal method [5] is an easy and non-destructive approach to measuring the thermal conductivity (κ) and interfacial thermal conductance (g) of material supported on a given substrate, wherein the temperature-dependent and power-dependent Raman responses of the material are measured. This technique has often been used for the measurement of κ and g of various 2D layered materials [1,5,6]. Therefore, we have used this technique for the measurement of thermal conductivity (κ) and interface thermal conductance (g) of $SnS_2$ flakes supported on $SiO_2$/Si substrates. The κ and g for the different thicknesses of $SnS_2$ supported on $SiO_2$/Si substrate were estimated by using the Raman spectroscopy-based measurement using the approach developed by Cai *et al.* [6]. A 2D layer when supported on a substrate can dissipate the heat through the in-plane and out-of-plane heat conduction channels (neglecting the contribution from convection and radiation losses). The temperature distribution in the sample can be obtained from the heat diffusion equation in cylindrical coordinates as follows [1,6]:

$$\frac{1}{r}\frac{d}{dr}\left(r\frac{dT}{dr}\right) - \frac{g}{\kappa l}(T - T_a) + \frac{q}{\kappa} = 0 \qquad (1)$$

Here $T$, $T_a$, $g$, $\kappa$, $l$, and $q$ represent the temperature at the position $r$, ambient temperature, interface conductance per unit area, thermal conductivity, the thickness of the sample, and volumetric heat coefficient, respectively. The volumetric optical heating q can be represented as:

$$q = \frac{q_0}{l} exp\left(-\frac{r^2}{r_0^2}\right) \qquad (2)$$

Here $q_0$ and $r_0$ represent the peak absorbed laser power per unit area at the center of the laser beam spot and the radius of the laser beam spot, respectively. The total absorbed laser power $Q$ is then

$$Q = \int_0^\infty q_0 \, exp\left(-\frac{r^2}{r_0^2}\right) 2\pi r dr = q_0 \pi r_0^2 \qquad (3)$$

Substituting $\theta = (T - T_a)$ and $z = \left(\frac{g}{\kappa l}\right)^{\frac{1}{2}} r$ into equation (1), we get a nonhomogeneous Bessel's equation:

$$\frac{d^2\theta}{dz^2} + \frac{1}{z}\frac{\partial \theta}{\partial z} - \theta = -\frac{q_0}{g} exp\left(-\frac{z^2}{z_0^2}\right) \qquad (4)$$

The solution to equation (4) is given as

$$\theta(z) = C_1 I_0(z) + C_2 K_0(z) + \theta_p(z) \qquad (5)$$

$I_0(z)$ and $K_0(z)$ are the zero-order modified Bessel functions of the first and second kind, respectively.

$$\theta_p(z) = I_0(z) \int K_0(z) \frac{\pi q_0}{2g} exp\left(-\frac{z^2}{z_0^2}\right) dz - K_0(z) \int I_0(z) \frac{\pi q_0}{2g} exp\left(-\frac{z^2}{z_0^2}\right) dz \qquad (6)$$

The boundary conditions $\left(\frac{d\theta}{dz}\right)_{z=0} = 0$ and $\theta(z \to \infty) = 0$ yield $C_2 = 0$ and $C_1 = -\left(\frac{\theta_p(z)}{I_0(z)}\right)$, which approach a constant value for large z.

The temperature rise in the sample measured by the Raman laser beam is given by:

$$\theta_m = \frac{\int_0^\infty \theta(r) exp\left(-\frac{r^2}{r_0^2}\right) rdr}{\int_0^\infty exp\left(-\frac{r^2}{r_0^2}\right) rdr} \qquad (7)$$

We define the measured thermal resistance as $R_m \equiv \frac{\theta_m}{Q}$. Based on equations (3) and (7), we get

$$R_m = \frac{\int_0^\infty \left(-I_0(z)\frac{\theta_p(z)}{I_0(z)} + \theta_p(z)\right) exp\left(-\frac{r^2}{r_0^2}\right) rdr}{\int_0^\infty exp\left(-\frac{r^2}{r_0^2}\right) rdr \int_0^\infty q_0 exp\left(-\frac{r^2}{r_0^2}\right) 2\pi rdr} \qquad (8)$$

*For measurements with 50× objective:*

$$R_{m(50\times)} = \frac{\int_0^\infty \left(-I_0(z)\frac{\theta_p(z)}{I_0(z)} + \theta_p(z)\right) exp\left(-\frac{r^2}{r_1^2}\right) r dr}{\int_0^\infty exp\left(-\frac{r^2}{r_1^2}\right) r dr \int_0^\infty q_1 exp\left(-\frac{r^2}{r_1^2}\right) 2\pi r dr} \quad (9)$$

*For measurement with 20× objective:*

$$R_{m(20\times)} = \frac{\int_0^\infty \left(-I_0(z)\frac{\theta_p(z)}{I_0(z)} + \theta_p(z)\right) exp\left(-\frac{r^2}{r_2^2}\right) r dr}{\int_0^\infty exp\left(-\frac{r^2}{r_2^2}\right) r dr \int_0^\infty q_2 exp\left(-\frac{r^2}{r_2^2}\right) 2\pi r dr} \quad (10)$$

To avoid the artificial shift in mode frequency, we have used $R_m = \frac{\partial \theta_m}{\partial Q}$ instead of $R_m = \frac{\theta_m}{Q}$.

To note that $R_m$ depends on $\kappa$ and g. The $R_m$ can be experimentally obtained using the following relation

$$R_m = \frac{\partial \theta_m}{\partial Q} = \frac{\partial \omega}{\partial Q}\frac{\partial \theta_m}{\partial \omega} = \chi_P(\chi_T)^{-1} \quad (11)$$

Where $\chi_P$ and $\chi_T$ are first-order power and temperature coefficients, [ $\chi_P = \left(\frac{d\omega}{dP}\right)_T$ or $=\left(\frac{\partial \omega}{\partial Q}\right)$ and $\chi_T = \left(\frac{d\omega}{dT}\right)_P$ or $=\left(\frac{\partial \omega}{\partial \theta_m}\right)$] respectively.

The $R_m$ values obtained experimentally using equation (11) from the temperature-dependent and power-dependent Raman data for the 50× and 20× objective lenses are used to avoid any calibration error in the $\kappa$ and g values for the samples by using equation 8. To solve it numerically, we have used $R_m = \frac{\partial \theta_m}{\partial Q}$ ratios for 50× and 20× to obtain the accurate results.

**Calculations of κ and g:**

(i) **For 1L (SnS₂ on SiO₂/Si):**

*In case of 50× objective*

$\chi_P$ = -0.090 cm⁻¹/mW

$\chi_T$ = -0.019 cm⁻¹/K

Therefore, $R_m = \frac{\partial \theta_m}{\partial Q} = \frac{\partial \omega}{\partial Q} \frac{\partial \theta_m}{\partial \omega} = \chi_P(\chi_T)^{-1} = 4.736$ K/mW

*In the case of 20× objective*

$\chi_P$ = -0.039 cm$^{-1}$/mW

$\chi_T$ = -0.016 cm$^{-1}$/K

Therefore, $R_m = \frac{\partial \theta_m}{\partial Q} = \frac{\partial \omega}{\partial Q} \frac{\partial \theta_m}{\partial \omega} = \chi_P(\chi_T)^{-1} = 2.437$ K/mW

Hence, $R_m$ ratio (of 50× to 20×) = (4.736/2.437) = 1.943

The peak absorbed laser power per unit area ($q_1$) corresponding to 50× objective: $q_1 = \frac{Q_1 \alpha}{\pi r_1^2}$

where,

$Q_1$ = Total absorbed laser power

$r_1$ = Laser spot radius for 50× objective = 0.64 μm

$\alpha$ = Optical absorption coefficient of monolayer SnS$_2$ = 0.0724 [see note S5]

$q_1 = \frac{1.38 \times 10^{-3} \times 0.0724}{\pi \times (0.64 \times 10^{-6})^2}$ Wm$^{-2}$ = 7.764 × 10$^7$ W m$^{-2}$

Similarly, for the 20× objective, the peak absorbed laser power per unit area ($q_2$): $q_2 = \frac{Q_2 \alpha}{\pi r_2^2}$

where,

$Q_2$ = Total absorbed laser power

$r_2$ = Laser spot radius for 20× objective = 0.94 μm

$\alpha$ = Optical absorption coefficient = 0.0724 [see note S5]

$q_2 = \frac{2.41 \times 10^{-3} \times 0.0724}{\pi \times (0.94 \times 10^{-6})^2}$ W m$^{-2}$ = 6.285 × 10$^7$ W m$^{-2}$

Thickness ($l$) of SnS$_2$ (monolayer) = 0.7 nm.

We have used the information of absorbed power or volumetric heat coefficient ($q_{1,2}$), laser spot radius ($r_{1,2}$), and thickness of the sample ($l$) in the equation (8) with the respective R$_m$ value of the corresponding objective (50× or 20×) and solved the equation (8) numerically to estimate κ and g by taking the ratio of $R_m$ for 50× to 20×. The estimated values for 1L SnS$_2$ (SnS$_2$ on SiO$_2$/Si) sample are κ = 3.2 W m$^{-1}$ K$^{-1}$ and g = 0.53 MW m$^{-2}$ K$^{-1}$

(ii) **For 2L (SnS$_2$ on SiO$_2$/Si):**

*In case of 50× objective*

$\chi_P$ = -0.084 cm$^{-1}$/mW

$\chi_T$ = -0.015 cm$^{-1}$/K

Therefore, $R_m = \frac{\partial \theta_m}{\partial Q} = \frac{\partial \omega}{\partial Q}\frac{\partial \theta_m}{\partial \omega} = \chi_P(\chi_T)^{-1}$ = 5.600 K/mW

*In case of 20× objective*

$\chi_P$ = -0.045 cm$^{-1}$/mW

$\chi_T$ = -0.017 cm$^{-1}$/K

Therefore, $R_m = \frac{\partial \theta_m}{\partial Q} = \frac{\partial \omega}{\partial Q}\frac{\partial \theta_m}{\partial \omega} = \chi_P(\chi_T)^{-1}$ = 2.647 K/mW

Hence, $R_m$ ratio (of 50× to 20×) = (5.600/2.647) = 2.115

The peak absorbed laser power per unit area (q$_1$) corresponding to 50× objective: $q_1 = \frac{Q_1 \alpha}{\pi r_1^2}$

where,

$Q_1$ = Total absorbed laser power

$r_1$ = Laser spot radius for 50× objective = 0.64 μm

$\alpha$ = Optical absorption coefficient of 2L SnS$_2$ = 0.0724×2 = 0.145 [see note S5]

$q_1 = \frac{1.38 \times 10^{-3} \times 0.145}{\pi \times (0.64 \times 10^{-6})^2}$ Wm$^{-2}$ = 1.555 × 10$^8$ W m$^{-2}$

Similarly, for the 20× objective, the peak absorbed laser power per unit area (q$_2$): $q_2 = \frac{Q_2 \alpha}{\pi r_2^2}$

where,

$Q_2$ = Total absorbed laser power

$r_2$ = Laser spot radius for 20× objective = 0.94 μm

$\alpha$ = Optical absorption coefficient = 0.145 [see note S5]

$q_2 = \frac{2.41 \times 10^{-3} \times 0.145}{\pi \times (0.94 \times 10^{-6})^2}$ W m$^{-2}$ = 1.259 × 10$^8$ W m$^{-2}$

Thickness ($l$) of 2L SnS$_2$ = 1.4 nm.

We have used the information of absorbed power or volumetric heat coefficient ($q_{1,2}$), laser spot radius ($r_{1,2}$), and thickness of the sample ($l$) in the equation (8) with the respective R$_m$ value of the corresponding objective (50× or 20×) and solved the equation (8) numerically to estimate κ and g by taking the ratio of R$_m$ for 50× to 20×. The estimated values for 2L SnS$_2$ (SnS$_2$ on SiO$_2$/Si) sample are κ = 4.6 W m$^{-1}$ K$^{-1}$ and g = 1.3 MW m$^{-2}$ K$^{-1}$

(iii) **For 5L (SnS$_2$ on SiO$_2$/Si):**

*In case of 50× objective*

$\chi_P$ = -0.045 cm$^{-1}$/mW

$\chi_T$ = -0.018 cm$^{-1}$/K

Therefore, $R_m = \frac{\partial \theta_m}{\partial Q} = \frac{\partial \omega}{\partial Q} \frac{\partial \theta_m}{\partial \omega} = \chi_P (\chi_T)^{-1}$ = 2.500 K/mW

*In the case of 20× objective*

$\chi_P$ = -0.020 cm$^{-1}$/mW

$\chi_T$ = -0.016 cm$^{-1}$/K

Therefore, $R_m = \frac{\partial \theta_m}{\partial Q} = \frac{\partial \omega}{\partial Q} \frac{\partial \theta_m}{\partial \omega} = \chi_P (\chi_T)^{-1}$ = 1.250 K/mW

Hence, R$_m$ ratio (of 50× to 20×) = (2.500/1.250) = 2.000

The peak absorbed laser power per unit area (q$_1$) corresponding to 50× objective: $q_1 = \frac{Q_1 \alpha}{\pi r_1^2}$

where,

$Q_1$ = Total absorbed laser power

$r_1$ = Laser spot radius for 50× objective = 0.64 μm

$\alpha$ = Optical absorption coefficient of 5L SnS$_2$ = 0.0724×5 = 0.362 [see note S5]

$q_1 = \frac{1.38 \times 10^{-3} \times 0.362}{\pi \times (0.64 \times 10^{-6})^2}$ Wm$^{-2}$ = 3.882 × 10$^8$ W m$^{-2}$

Similarly, for the 20× objective, the peak absorbed laser power per unit area (q$_2$): $q_2 = \frac{Q_2 \alpha}{\pi r_2^2}$

where,

$Q_2$ = Total absorbed laser power

$r_2$ = Laser spot radius for 20× objective = 0.94 μm

$\alpha$ = Optical absorption coefficient = 0.362 [see note S5]

$q_2 = \frac{2.41 \times 10^{-3} \times 0.362}{\pi \times (0.94 \times 10^{-6})^2}$ W m$^{-2}$ = 3.142 × 10$^8$ W m$^{-2}$

Thickness ($l$) of 2L SnS$_2$ = 3.2 nm.

We have used the information of absorbed power or volumetric heat coefficient ($q_{1,2}$), laser spot radius ($r_{1,2}$), and thickness of the sample ($l$) in the equation (8) with the respective R$_m$ value of the corresponding objective (50× or 20×) and solved the equation (8) numerically to estimate κ and g by taking the ratio of R$_m$ for 50× to 20×. The estimated values for 5L SnS$_2$ (SnS$_2$ on SiO$_2$/Si) sample are κ = 5 W m$^{-1}$ K$^{-1}$ and g = 3.6 MW m$^{-2}$ K$^{-1}$

**Supplemental Note S6: Absorption coefficient of 2H-SnS$_2$**

We have first exfoliated thin layers of SnS$_2$ from bulk single crystal onto the SiO$_2$/Si substrate to perform this measurement. We have recorded the Raman signal using the LABRAM HR Evolution Raman spectrometer (1800 lines/mm), fitted with a microscope with 50× (NA=0.5) using 532 nm laser with less than 1 mW laser power to minimize the heating effect on the sample. We have measured the absorption coefficient of SnS$_2$ thin flakes from the Raman and AFM measurements using the simple light attenuation model [7]. Let us consider an n-layer $SnS_2$ flake placed on a Si ($SiO_2$/Si) substrate. Therefore, the intensity of the Si Raman mode would decrease, as compared to that of bare Si substrate, as the number of layers (n) of $SnS_2$ flake increases because of the absorption coefficient associated with each layer. Therefore, to determine the absorption coefficient of $SnS_2$ we have utilized the change intensity of the Si Raman peak induced by the varying thickness of $SnS_2$ as shown in Table S1 and Fig S5. The thickness of 1L, 2L, and 5L are confirmed by AFM as shown in the main text Fig 1(b) and the thickness of 6L and 7L are shown in Fig S3. From this data, we have measured that 7.24 % visible light is absorbed through one monolayer $SnS_2$ which is close to the absorption coefficient of the $MoS_2$ monolayer [8]. The attenuation of light can be expressed as [7]:

$I = I_0 e^{-2\alpha t}$

Where I represents the intensity of the silicon Raman peak with an $SnS_2$ flake on it, $I_0$ is the intensity of the silicon Raman peak of the bare substrate, t is the flake thickness and α is the absorption coefficient in $cm^{-1}$ unit.

Table S1: Characterization of the $SnS_2$ flakes from AFM and Raman measurements and calculated absorption coefficient for each flake.

| Spot | Si Raman mode (cm$^{-1}$) | Si Intensity (Area under the curve) | Si Intensity Ratio($I/I_0$) | Ln($I/I_0$) | Absorption coefficient ($\times 10^6$) | Flake thickness (nm) | Absorption coefficient in percentage ($\times 100$) | No of layer of SnS$_2$ |
|---|---|---|---|---|---|---|---|---|
| A | 520.6 | 60932 ($I_0$) | | | | | | 0 (Bare Si) |
| B | 520.5 | 51194 | 0.84018 | -0.174 | 1.210 | 0.70 | 0.085 | 1 |
| C | 520.5 | 53836 | 0.88354 | -0.124 | 0.876 | 0.70 | 0.061 | 1 |
| D | 520.6 | 52774 | 0.86612 | -0.144 | 1.021 | 0.70 | 0.071 | 1 |
| E | 520.6 | 47729 | 0.78333 | -0.244 | 0.380 | 3.20 | 0.121 | 5 |
| F | 520.6 | 42126 | 0.69136 | -0.369 | 0.479 | 3.85 | 0.184 | 6 |
| G | 520.7 | 44083 | 0.72348 | -0.324 | 0.358 | 4.50 | 0.161 | 7 |

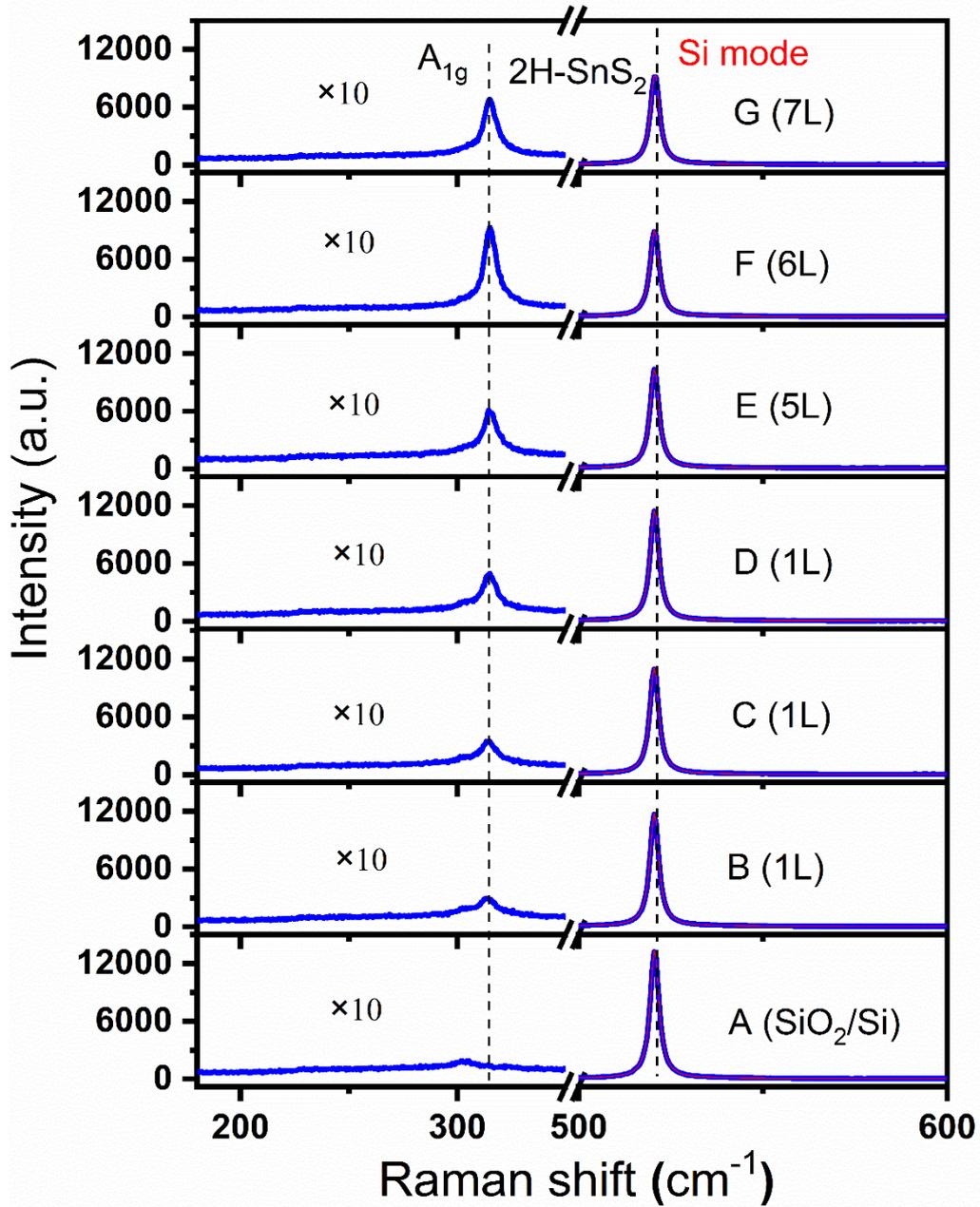

**FIG. S5**. Raman spectra of Si substrate with and without the 2H-SnS$_2$ layers show the variations in intensity. The left part of the spectra up to 350 cm$^{-1}$ has been magnified by 10 times for clarity.

**Supplemental Note S7: Dependence of thermal conductivity (κ) and interfacial thermal conductance per unit area (g) on the laser spot size**

In addition to using 50× and 20× objective lenses, as discussed above, we have performed the temperature and laser-dependent Raman measurement using a 100× objective lens to investigate the dependence of thermal conductivity on the laser spot radius. First, we have performed a thorough Raman study on the same 1L, 2L, and 5L flakes by varying the temperature from 300 K to 420 K at a fixed laser power below 1mW to reduce the locale heating of the sample. Then, we performed a similar measurement by varying the laser power using a neutral density filter up to ~ 6 mW laser power at room temperature for 50× and 20× objective lenses. On the contrary, we have worked in the power range of 0.1-1.5 mW for 100× objective because at laser powers higher than this can potentially damage the flake due to higher power density for 100× objective. We have observed red-shift of the $A_{1g}$ mode with the increase of temperature and the laser power due to the softening of the phonon modes, as can be seen in the fig S6.

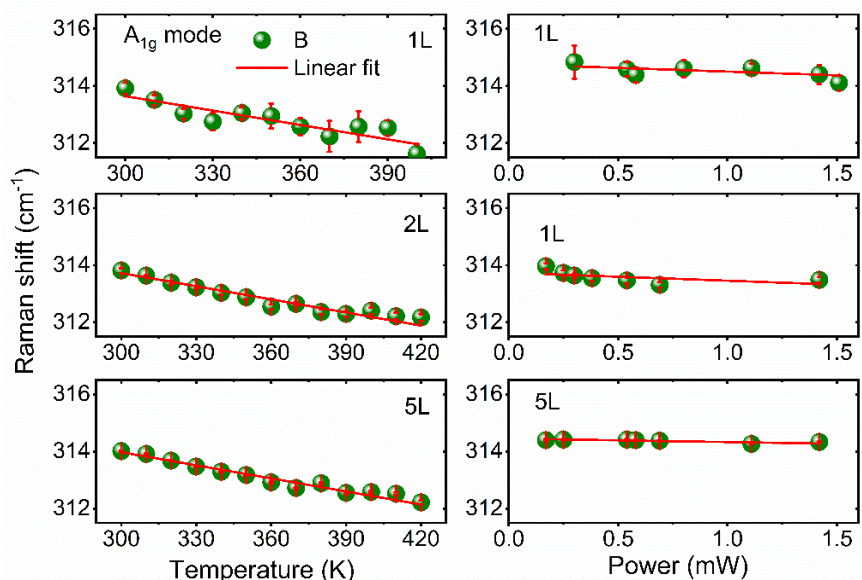

**FIG. S6.** Temperature and power dependence of $A_{1g}$ Raman modes for 1L, 2L, and 5L $SnS_2$ flakes. The slopes give the first-order temperature and power coefficients with 100× lens, which are enlisted in Table S2.

*Table S2: First-order temperature and power coefficients of the various investigated flakes.*

| Sample | Modes | Objective | $\chi_T$ (cm$^{-1}$/K) | $\chi_P$ (cm$^{-1}$/mW) |
|---|---|---|---|---|
| **1L** | $A_{1g}$ | 100× | -0.017 | -0.254 |
| **2L** | $A_{1g}$ | 100× | -0.015 | -0.270 |
| **5L** | $A_{1g}$ | 100× | -0.015 | -0.120 |

The 100× lens data have been used in combination with 50× data given in main text table S1 to estimate the values of in-plane thermal conductivity ($\kappa$) and interfacial thermal conductance per unit area with the substrate (g)

*Table S3: Measured $\kappa$ and g value by using the 100× and 50× objective lenses.*

| Samples | Thermal conductivity ($\kappa$) in W m$^{-1}$ K$^{-1}$ | Interfacial thermal conductance per unit area (g) in MW m$^{-2}$ K$^{-1}$ |
|---|---|---|
| **1L** | (3.1-3.2) | 0.53 |
| **2L** | (4.3-4.6) | 1.30 |
| **5L** | (5-5.2) | 3.6 |

It should be noted that the estimated values, given above, are comparable to the estimates obtained using the set of 50× and 20× objectives. This observation supports to the fact that since the lateral heating length (~ 65 nm) is smaller than the laser spot size (~ 300-900 nm used for our measurements) then the measurement of thermal conductivity becomes insensitive to the spot size, as also shown by Yalon *et al*. [9]. The estimation of the lateral heating length is discussed below.

To obtain these two unknown parameters κ and g, we have used the combination of two objective lenses 50× and 20× (with spot size ~ 0.64 and ~ 0.94 μm, respectively) and the results were confirmed by an additional measurement with the help of a 100× objective lens (with spot radius ~ 0.34 μm) in combination with 50× objective lens (spot size ~ 0.64 μm). Based on these measurements, we observe that the κ and g are insensitive to the laser spot-size and laser power (at least within the range used in these measurements to avoid any possible high-power-induce degradation of the 2D layers). To recall that for 2D layers, the lateral heating length ($L_h$) is a characteristic length up to which the heat flows in the plane before it sinks. Yalon *et al*. [9] have shown by simulation that if the $L_h$ is much shorter than the laser spot-size (which is ~ 300-900 nm in our case) then the measurement of thermal conductivity becomes insensitive to the spot-size. For the monolayer film of SnS$_2$, the lateral heating length $L_h = \left(\frac{\kappa_{1L} t_{1L}}{g}\right)^{\frac{1}{2}} \approx 65$ nm ( see the reference [9]), where, $\kappa_{1L}$= 3.2 W m$^{-1}$ K$^{-1}$ and $t_{1L}$= 0.7 nm represent the thermal conductivity and thickness of the 1L-SnS$_2$, and $g = 0.53$ W m$^{-2}$ K$^{-1}$ is the interfacial thermal conductance per unit area between the monolayer SnS$_2$ and the SiO$_2$/Si substrate. On the other hand, for the laser power, there is a limitation that, above a certain laser power (above 6 mW for 50× and 20× lenses and ~ 1.5 mW for 100× objective) the 2D layers may get damaged. So, we have avoided using higher laser power to check sensitivity of the measurement.

**Supplemental Note S8: Error calculations**

The errors in the estimation of the thermal conductivity were calculated by the root-sum-square-error propagation approach.

$$\text{Error} = \sqrt{\left(\frac{\Delta r_{01}}{r_{01}}\right)^2 + \left(\frac{\Delta r_{02}}{r_{02}}\right)^2 + \left(\frac{\Delta \alpha}{\alpha}\right)^2 + \left(\frac{\Delta p}{p}\right)^2 + \left(\frac{\Delta T}{T}\right)^2}$$

$$= \sqrt{\left(\frac{0.01}{0.94}\right)^2 + \left(\frac{0.01}{0.64}\right)^2 + \left(\frac{0.14}{0.78}\right)^2 + \left(\frac{0.001}{1.1}\right)^2 + \left(\frac{5}{300}\right)^2}$$

$= 0.1806$ or $18.06\%$

Where, $r_{01}$ is the spot radius for 20× objective, $r_{02}$ is spot radius for 50× objective, α is absorption coefficient, p is the absorbed laser power, and T is the temperature.

For the 1L sample

Error in κ is $3.2 \times 0.18 = 0.57$ W m$^{-1}$ K$^{-1}$

In g is $0.53 \times 0.18 = 0.09$ W m$^{-2}$ K$^{-1}$

Similarly, for 2L

Error in κ is 0.82 W m$^{-1}$ K$^{-1}$

In g is 0.23 W m$^{-2}$ K$^{-1}$

For 5L,

Error in κ is 0.90 W m$^{-1}$ K$^{-1}$

In g is 0.65 W m$^{-2}$ K$^{-1}$

Similarly, when we take spot radius for 50× and 100× objectives then the error comes out to be 0.1809 or 18.09 % which is comparable to the error discussed above for measurements with 50× and 20× objectives.

**Supplemental Note S9: Heat capacity calculation**

FIG. S7 represents the calculated heat capacity of mono-layer $SnS_2$ using density functional theory. At room temperature, the heat capacity is below 32 joule/ k-mole, which is very less as compared to other 2D materials.

**FIG. S7:** Variation of heat capacity of $SnS_2$ monolayer with respect to temperature.

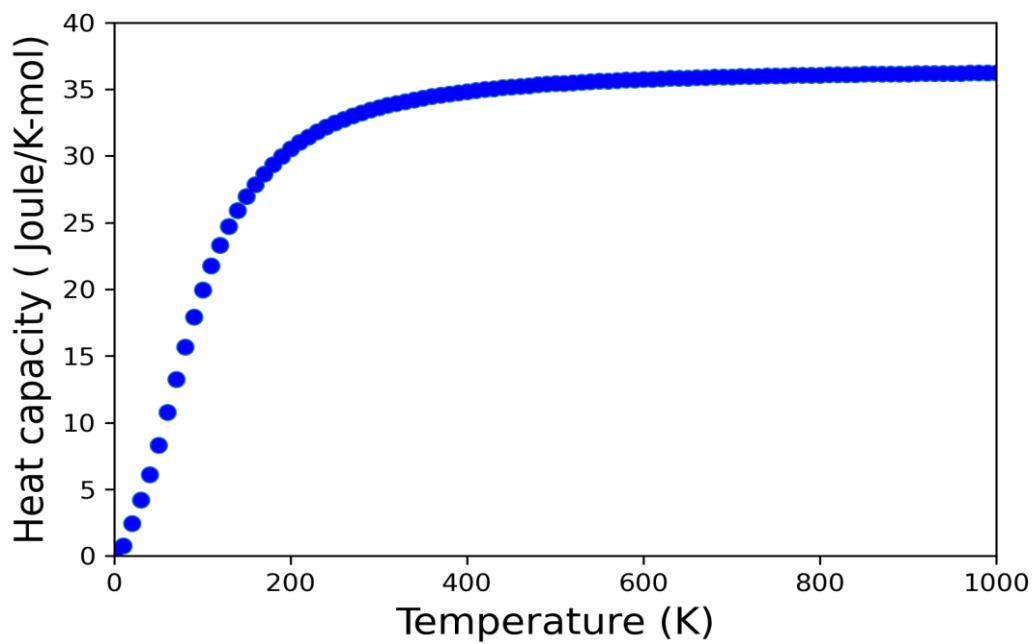